# Проектирование и реализация комплексной информационной системы поддержки научных исследований

Д.Е. Прокудин

Санкт-Петербургский государственный университет
hogben.young@gmail.com

**Аннотация**

Информатизация научно-исследовательской деятельности привела к созданию большого специализированных информационных ресурсов, платформ, сервисов и программного обеспечения для поддержки научных исследований. Однако присущие им недостатки не позволяют в полной мере реализовать комплексную поддержку научной деятельности, а отсутствие единой точки входа разбивают научное сообщество на раздробленные группы «по интересам». В статье на основе анализа особенностей существующих решений и подходов к обеспечению средствами информационно-коммуникационных технологий различных видов научной деятельности и с учётом структуры научно-исследовательской деятельности сформулированы и предложены основные принципы проектирования и реализация комплексной информационной системы поддержки научных исследований.

## 1. Проблемы организации научной деятельности в условиях развивающегося информационного общества

В условиях непрерывно развивающегося информационного общества процессы информатизации кардинальным образом перестраивают любую сферу человеческой деятельности. Через внедрение информационно-коммуникационных технологий претерпевает изменения и научная деятельность. Процессы глобализации ведут к построению единого информационного пространства научных исследований. Создаётся всё больше специализированных информационных ресурсов, платформ, сервисов и программного обеспечения для поддержки научных исследований [3].

На современном этапе наиболее развитыми являются технологии представления научных публикаций в цифровой форме с предоставлением удобных механизмов поиска: электронные каталоги библиотек, полнотекстовые базы научной информации, реферативные базы научной периодики, цифровые коллекции и репозитарии научных текстов, электронные издательские платформы, различные электронные периодические научные издания и т.п. Что позволяет в полной мере охватить информационно-поисковый вид научной деятельности. Однако, далеко не все источники являются доступными всему научному сообществу – в часть из них доступ осуществляется по подписке, на которую могут потратиться далеко не все вузы или иные научные учреждения (а что при этом говорить о доступе не аффилированных исследователей); некоторые ресурсы являются внутренними корпоративными хранилищами информации и доступны только аффилированным сотрудникам, т.е., по большому счёту, эти ресурсы составляют так называемое пространство Deep Web, доступный далеко не каждому страждущему учёному и и исследователю.

Для обнародования результатов научных исследований доступны такие технологии как: сервисы эдутех (edutech), социальные медиа сервисы, виртуальные электронные научные конференции, различные цифровые репозитарии научных текстов. Необходимо отметить, что при этом почти вся ответственность за соблюдение научной и публикационной этики ложиться на того, кто публикует информацию. Зачастую нет уверенности в достоверности не только публикуемых данных (так как отсутствует обязательное научное рецензирование или оно сведено к простой формальности), но и в биографических сведениях профиля того, от чьего имени обнародуется эта информация.

В области научной коммуникации используются как сервисы вербального непосредственного (синхронного) взаимодействия – Skype (http://skype.com), Oovoo (http://oovoo.com) и пр.; платформы для проведения вебинаров и телеконференций, так и опосредованного (асинхронного) общения – электронная почта; социальные сервисы в сети Интернет и т.п. Такие сервисы являются массовыми и при наличии каналов связи с достаточной пропускной способностью и невысоким сетевым трафиком позволяют осуществлять оперативную научную коммуникацию. Но некоторые из этих решений являются высокотехнологичными и дорогостоящими, что не позволяет в полной мере использовать их для повседневной массовой коммуникации в научной среде.

Помимо этого существует ещё один вид специализированного программного обеспечения, востребованного в научной среде - это библиографические менеджеры. На современном этапе развития технологий они помимо решения задачи организации библиографии, её классификации и упорядочения, также позволяют выполнять ряд других функций: автоматическое формирование библиографических описаний со страниц сетевых ресурсов, вставка в текстовом редакторе библиографических ссылок и автоматическое формирование списков использованной литературы, организация совместной работы над составлением библиографии и пр. Это возможно благодаря наличию динамически подключаемых к обозревателям сети и текстовым редакторам программных модулей, а также применению облачных технологий и элементов социальных сетей. Однако все преимущества от применения сетевых сервисов нивелируются ограниченностью бесплатного тарифного плана.

К основным недостаткам существующих решений можно отнести следующие:
– невозможность охвата всех видов научной деятельности в рамках одного решения;
– как следствие, использование множества учётных записей на различных сетевых сервисах, что неудобно как для пользователя, так и для тех, с кем он коммуницирует;
– ограниченный доступ к некоторым сетевым сервисам и информационным системам (наличие подписки, платные услуги и т.д.).

Эти и другие недостатки не позволяют в полной мере реализовать комплексную поддержку научной деятельности, а отсутствие единой точки входа, т.е. отсутствие единой учётной записи разбивают научное сообщество на раздробленные группы «по интересам».

## 2. Информационные системы для научных исследований

Среди большого числа различных по своему назначению сетевых сервисов, информационных систем, программных платформ и программных продуктов можно выделить наиболее популярные в научно-исследовательской среде, которые позволяют решать ряд задач, решаемых при проведении научных исследований. Среди них особенно выделяются комплексные системы из разряда образовательных технологий (edutech, эдутех), которые созданы на стыке медиа и сервисной составляющей. Их объединяют как электронные библиотеки учебного и научного контента, так и медиа вокруг научной тематики – видеохостинги, блог-платформы, системы коллективной работы исследователей. К ним можно отнести также системы, в которых смыкаются технологии как специализированной социальной сети, так и обычного файлового хранилища для контента, наполняемого специалистами (кураторами, модераторами и так далее). Как правило, такие проекты организовывают или сами преподаватели дисциплин в вузах (которые активно принимают участие в научных исследованиях), или связанные с издательским или медиабизнесом предприниматели. В США такие инициативы исходят преимущественно от коммерческих вузов и издательских домов, в Европе – от общественных организаций. Россия пока находится в процессе формирования предложения, но движется в сторону американского подхода [5]. Возникая зачастую «снизу», как потребность членов научного сообщества такие проекты для своего дальнейшего существования и развития быстро переходят к коммерциализации. И в качестве коммерческих проектов они ориентированы на использование одной из следующих бизнес-моделей:
– продавать данные своих посетителей или подписчиков рекламодателям, потому что таргетированная реклама в платежеспособном сегменте западных учёных стоит дороже обычной (ResearchGate);
– делать аналитику по загруженному контенту и подбирать необходимые исследования за отдельную плату (Academia.edu);
– предоставлять дополнительные платные сервисы вроде хранилища материалов и организации дискуссионных площадок для вузов (Mendeley) [5].

Рассматривая же потребности самих представителей научной среды, можно привести основные возможности данных систем.

Так, например, Academia.edu и ResearchGate позволяют зарегистрированным пользователям:
- загружать в систему свои научные тексты, которые будут доступны для других пользователей. При этом в новостной ленте пользователей уведомления об этих статьях будут появляться либо при совпадении ключевых слов статей и интересов пользователей, либо от пользователей, с которыми установлены «академические» контакты в самом сервисе. Это способствует распространению научного знания (обнародование результатов научных исследований);
- пользоваться аналитическими возможностями систем: иметь доступ к статистике просмотра своего профиля, публикаций и загрузки каждой публикации. При этом могут быть определены страна происхождения запроса, сам поисковый запрос и внешние ресурсы, с которых произошёл заход и пр.;
- производить поиск по научным интересам пользователей и устанавливать с ними академические контакты.

Помимо этого, например, сервис Academia.edu (http://academia.edu) позволяет регистрировать в системе профили сайтов научных журналов или электронных их версий и отображать в новостной ленте пользователей новости этих журналов (например, выход в свет номеров или публикацию статей), которые берутся из RSS-ленты соответствующего издания. Каждый пользователь может заполнить свои анкетные данные, которые могут быть использованы для контекстных предложений на вакантные позиции в различные учреждения. Для наиболее полного представления информации о пользователе есть возможность представить ссылки на внешние ресурсы: профили наиболее популярных сетей, профиль Академии Google или иные произвольные ссылки.

Однако, сервис Academia.edu не рассчитан на комплексную поддержку проведения научных исследований, а пользователи, в основном, помещают в системе уже опубликованные статьи.

В отличие от него сервис ReserchGate (http://researchgate.net) более развит в различных средствах. Так, например, по тематике научных интересов пользователя ему может быть предложено ответить на какой-нибудь вопрос, заданный другим пользователем, т.е. выступить в качестве научного эксперта. Также можно выступить в качестве рецензента публикаций, размещённых в системе (поиск производится по названиям представленных статей). Помимо этого предоставляется возможность поиска цитирования своих статей в системе и отображения информации о цитировании. Неоспоримым преимуществом сервиса ReserchGate является наличие организации проектов с привлечением к совместной работе над ними других пользователей системы, с которыми установлены академические контакты. Эта возможность реализована через создание именованного пространства, представляющего собой защищённый контейнер, в котором можно создавать опыты, тесты, стенды (bench) и размещать в них различные файлы и проводить обсуждения с другими участниками проекта в виде комментариев и реплик, организованных в офф-лайн чат. При этом также, как и в Academia.edu, в сервисе ReserchGate можно публиковать свои статьи и получать статистику по их просмотру, скачиванию и цитированию. Ещё одной интересной особенностью сервиса является возможность для каждого участника набора очков за размещение публикаций, установление академических контактов и ответов на предлагаемые по тематике своих научных интересов вопросы. С одной стороны, наличие такой количественной оценки активности пользователей является «новым способом измерения научной репутации» (как отмечено на соответствующей странице системы), с другой – является стимулирующим механизмом активизации деятельности пользователей сервиса.

Более узкоспециализированным является сервис Mendeley (http://mendeley.com), который рассчитан, прежде всего, на работу с библиографией. Основным инструментом является одноимённый программный продукт (свободно распространяемый), который позволяет на персональном компьютере пользователя организовать упорядоченное хранилище различных тематических библиографий, производить по ним эффективный поиск и получать оперативный доступ к соответствующим ресурсам – файлам, хранящимся в локальной системе, или ресурсам сети Интернет. Наличие программных модулей для интеграции ПО Mendeley с различными Интернет-обозревателями и популярными текстовыми редакторами позволяет расширить возможности рассматриваемого ПО и использовать его как для оперативного импорта в локальное хранилище найденной информации в сети Интернет, так и вставлять в свои тексты библиографические ссылки, а также автоматически формировать пристатейные списки использованной литературы. При этом есть возможность выбирать различные стили цитирования и представления списков литературы (в том числе и по ГОСТу). Наличие облачного сервиса Mendeley расширяет возможности пользователей и позволяет им синхронизировать свою локальную библиографию с удалённым хранилищем, что даёт возможность оперативного доступа к ней с любого устройства, подключенного к сети Интернет, а также позволяет осуществлять коллективную работу над общей библиографией для других пользователей сервиса. Правда, эти возможности в базовом (бесплатном варианте) очень ограничены и за их расширение необходимо платить абонентскую плату, что накладывает известные ограничения на использование этого инструмента в академической среде. Указанные возможности рассчитаны, скорее всего, не на отдельных пользователей, а на использование сервиса Mendeley организациями для своих сотрудников.

К информационным системам и сервисам, рассчитанным на поддержку научных исследований можно также отнести проект ORCID (http://orcid.org), который является общедоступным и реализует универсальный идентификатор учёного. Этот проект создан при поддержке ведущих мировых университетов, научных обществ, научных издательств и других организаций, ориентированных на научную деятельность или её поддержку. В своём профиле пользователь может заполнить биографическую информацию, информацию об обучении, об опыте работы; информацию об участии в выполнении грантовых исследований, а также список своих публикаций, которые могут быть получены автоматически из основных баз данных научных публикаций и реферативных баз (например, CrossRef, PubMed, ResearcherID, Scopus и др.), что говорит об интеграции сервиса ORCID с внешними информационными системами. Определение в своём профиле списка ключевых слов своих научных интересов является основой механизма поиска персоналий из реестра сервиса и установления академических контактов. Наличие в ORCID сервиса для организаций позволяет:

– для фондов вести учёт грантов и публиковать информацию о них;
– для научно исследовательских организаций анализировать информацию о сотрудниках реестра;

- для издательств управлять базой данных авторов и производить поиск потенциальных авторов и т.п.

Этот проект интересен тем, что на глобальном уровне можно составить единый реестр учёных, исследователей и научных организаций, а также использовать уникальный идентификатор для получения полной и достоверной информации о личности учёного.

Говоря об отечественной науке, можно упомянуть недавно запущенную под эгидой министерства образования и науки РФ систему «Карта российской науки» (http://mapofscience.ru). Как отмечается в пресс-релизе к разработке этой системы, создание информационно-аналитической системы «Карта российской науки» имеет целью интеграцию российской науки в мировую, которая реализует общемировые тенденции открытия массивов научных данных, создания электронных научных справочников, создания специализированных научных сетей и сервисов для совместной работы исследователей. «Кроме того, проект позволит обеспечить более эффективное взаимодействие между научными учреждениями, исследователями и хозяйствующими субъектами». [2] Исходя из озвученных на прошедшем в Министерстве образования и науки РФ совещании по вопросам текущего статуса и развития проекта «Карты российской науки» задач создания проекта можно сделать вывод, что такая система, скорее всего, будет являться некоторым инструментом мониторинга и контроля научной деятельности в нашей стране. Находящаяся на сегодняшний день в тестовой эксплуатации информационно-аналитическая система содержит, например, информацию о публикациях учёных на основании выгрузки данных из РИНЦ (Российский индекс научного цитирования, созданный на базе Научной электронной библиотеки http://elibrary.ru) и реферативной базы публикаций, работающей на платформе *Web of Science* компании Thomson Reuters (*https://webofknowledge.com*), которые в принципе не являются полными и не отражают объективную картину публикационной активности российских учёных.

Ещё одним примером отечественной системы является информационно-аналитическая система сопровождения научно-исследовательской деятельности Санкт-Петербургского государственного университета (https://ias.csr.spbu.ru/). Эта система является служебной, т.е. полноценно с ней могут работать только аффилированные с СПбГУ сотрудники, для которых есть возможность вести список своих проектов (поддержанных или не поддержанных внешними научными фондами); НИР, проводимых в рамках деятельности в СПбГУ и список своих публикаций. Помимо этого, пользователям доступна информация о проектах, НИР и публикациях других пользователей; информация об имеющихся в СПбГУ ресурсных центрах и научном оборудовании; информация о периодических и продолжающихся изданиях, а также издательствах. Однако, как видится, единственной пользой (помимо получения справочной информации о научном потенциале СПбГУ и различных объявлений о конкурсах и грантах) для сотрудников является возможность формирования списка своих публикаций в формате ***doc***, а также подачи заявки на установление стимулирующей доплаты за научные публикации. Скорее всего эта информационно-аналитическая система является корпоративным средством аккумуляции и мониторинга информации о научном потенциале СПбГУ.

Существуют и другие сетевые информационные системы и сервисы, целевой аудиторией которых является научное сообщество [4, 6]. Однако, у них достаточно ограниченные возможности и они не могут рассматриваться в качестве комплексных систем поддержки научной деятельности.

## 3. Принципы проектирования и реализации комплексной информационной системы поддержки научных исследований

Существуют различные подходы к определению структуры научно-исследовательской, «жизненного цикла исследования» [1, 7, 8, 9, 10]. Во многом они нацелены на формирование так называемого «брэнда» учёного, что как раз таки и реализуется в упомянутых нами основных интернет-сервисах (Academia.edu, ReserchGate) через возможность выдачи вакансий по научным интересам пользователей, а также просмотр профилей пользователей потенциальными работодателями, фондами, руководителями научно-исследовательских коллективов. Анализ этих подходов, а также анализ институционализации научно-исследовательской деятельности позволяет выделить основные её виды, регламентирующие её структуру и являющиеся инвариантными по отношению к предметной области, области знания, содержанию, методам и подходам конкретного научного исследования. К ним можно отнести:

- информационно-поисковый вид деятельности, результатом которого является составление библиографии по теме исследования;
- «констатирующий», на котором в результате научно-исследовательской деятельности рождается новое научное знание, воплощённое в научный текст;
- обнародование результатов научной деятельности в виде публикаций или выступлений с докладами на различных научных мероприятиях;
- научная коммуникация, дающая, с одной стороны, возможность получения «обратной связи» исследователю, а, с другой – служащая проведению совместных, коллективных исследований (что ёмко определяется понятием collaboration (анг. – сотрудничество).

Поэтому, как видится, основным принципом, учитываемым при проектировании комплексной информационной системы поддержки научных исследований является то, что такая система должна в полном объёме охватывать все эти основные инвариантные виды научной деятельности, реализовывая полный жизненный цикл научного исследования.

Исходя из этого основного принципа и логики институционализации организации научной деятельности можно выделить основные объекты проектируемой комплексной информационной системы:

1. Пользователь со всеми сопутствующими атрибутами (анкетные данные, аффилиация, ключевые слова научных интересов и т.д.).
2. Организации, с которыми аффилированы пользователи.
3. Организации и фонды, предоставляющие грантовую поддержку научных исследований.
4. Издательства, в которых издаются научные периодические и непериодические издания.
5. Научные периодические издания.
6. Научные сообщества, объединяющие пользователей (аналоги групп и «кругов» в социальных сетях) с возможностью вести дискуссию (аналоги чатов и форумов).
7. Библиографии, которые могут создаваться как персонально пользователями, так и сообществом пользователей.
8. Публикации пользователей, которые могут быть привязаны к издательствам и периодическим научным изданиям.
9. Проекты, которые могут объединять пользователей, библиографии, общие файлы и документы (тестовые, электронные таблицы, презентации и т.п.), а также могут быть привязаны к грантам научных фондов и организаций. Также с проектами могут быть ассоциированы публикации, выполнявшиеся в ходе их реализации.

Необходимо отметить, что все эти (а также могущие возникнуть в будущем) объекты должны иметь возможность установления взаимосвязей друг с другом, что позволит не только гибко и комплексно решать задачи организации совместной научной деятельности коллективами учёных, но и получать различные отчёты, например, списки проектов научного фонда, списки публикаций по проекту, списки публикаций издательства, статистические отчёты о научной деятельности аффилированных с организацией сотрудников и т.д. и т.п. Так что такая комплексная система может быть универсальным инструментом не только для реализации задач поддержки научных исследований, но и быть полезной для анализа деятельности научных организаций, фондов и издательств.

Ещё одним важнейшим принципом построения системы является выбор модели аутентификации пользователя. Нет необходимости создавать ещё одну учётную запись, которая только увеличит количество профилей пользователя в различных сетевых сервисах. Для аутентификации предлагается выбрать универсальный идентификатор учёного, реализованный в открытом проекте ORCID, что также позволит однозначно идентифицировать личность исследователя. А возможность интеграции с этой системой позволит исключить дублирования и искажения информации о личности учёного: анкетные данные, публикации, аффилиация, участие в проектах и т.д., которые можно импортировать и экспортировать между системами.

Разработка любой информационной системы, тем более такой комплексной, нацеленной на решение комплекса различных задач, должна идти по модульному принципу, т.е. при реализации каждой отдельной задачи она решается в рамках создания отдельного программного модуля, который в любой момент можно подключить к системе, наращивая тем самым её функциональные возможности (так, например, в качестве модуля можно реализовать сервис, который позволит предлагать пользователям на рецензирование препринты статей, размещённых в системе, что позволит осуществить концепцию открытого «свободного рецензирования» в рамках всего научного сообщества). При таком подходе нет необходимости запускать систему после окончательной разработки всех составляющих её сервисов и компонентов – она уже может быть введена в эксплуатацию с минимальным набором функционала постепенно развиваясь при этом. К тому же модульное построение системы позволяет более оперативно выполнять задачи отладки и тестирования как отдельных модулей, так и всей системы в целом.

## 4. Заключение

Предлагаемые принципы проектирования комплексной информационной системы поддержки научных исследований обсуждались с авторами недавно запущенного проекта ScienceList (http://sciencelist.ru), поэтому хотелось бы надеяться, что они будут учтены при дальнейшем развитии сервисов этого стартапа.

Но, что особенно важно, сквозным принципом, идеологией функционирования и развития информационных систем поддержки научных исследований должна быть открытость. Притом открытость как со стороны пользователей, организаций, издательств и фондов, так и со стороны разработчиков, которые могут быть вовлечены в развитие системы, создавая новые сервисы и модули исходя из тенденций развития научных исследований и запросов научного сообщества.

Поэтому такая комплексная информационная система должна быть реализована как Open Source проект, а её сопровождение должно быть построено на добровольных пожертвованиях. Иначе коммерциализация такого проекта рано или поздно приведёт к возникновению платных услуг и сервисов и, следовательно, перестанет быть одинаково доступным инструментом организации научной деятельности для любого учёного или исследователя. Поэтому основным принципом эксплуатации и развития комплексной информационной системы поддержки научных исследований должно быть соблюдение концепции организации научной деятельности как «общественного блага».

# Design and Implementation of an Integrated Information System to Support Scientific Research

Dmitry E. Prokudin

Computerization of research activities led to the creation of large specialized information resources, platforms, services and software to support scientific research. However, their shortcomings do not allow to fully realizing the comprehensive support of scientific activity, and the absence of a single entry point to divide the scientific community fragmented «groups' interests». The article based on analysing the existing solutions and approaches to the tools of information and communication technologies of various types of scientific activity, and taking into account the research lifecycle proposed and formulated the basic principles of designing and implementing an integrated information system to support scientific research.